\documentclass[10pt,journal,compsoc]{IEEEtran}

\usepackage{fancyhdr}

\RequirePackage[hyphens]{url}
\usepackage{cite}
\usepackage{amsmath,amssymb,amsfonts,bm}

\usepackage[linesnumbered,ruled,norelsize,lined]{algorithm2e}
\usepackage{algpseudocode}
\def\HiLi{\leavevmode\rlap{\hbox to \hsize{\color{gray!35}\leaders\hrule height .8\baselineskip depth .5ex\hfill}}}

\DontPrintSemicolon
\SetAlFnt{\small}

\SetCommentSty{algocommentstyle}
\SetKwComment{Comment}{$\triangleright \,$}{}

\usepackage{graphicx}
\usepackage{listings}
\usepackage{float} %
\newfloat{lstfloat}{htbp}{lop}
\floatname{lstfloat}{Listing}
\usepackage{gensymb}
\usepackage{textcomp}
\usepackage{xspace}
\usepackage{upgreek}
\def\BibTeX{{\rm B\kern-.05em{\sc i\kern-.025em b}\kern-.08em
		T\kern-.1667em\lower.7ex\hbox{E}\kern-.125emX}}
\usepackage[table,xcdraw]{xcolor}
\usepackage{hyperref}
\usepackage{booktabs}
\usepackage[flushleft]{threeparttable}
\usepackage{comment}
\usepackage{cleveref}
\usepackage[numbers]{natbib}
\usepackage{ifthen}
\usepackage{enumitem}
\usepackage{soul}
\usepackage{multirow}
\usepackage{rotating}
\newcommand*\rot{\rotatebox{90}}
\usepackage{makecell}
\usepackage{balance}
\usepackage{fancyvrb}
\usepackage{tcolorbox}

\usepackage{lipsum}  

\usepackage{tikz}
\usetikzlibrary{shapes,arrows}

\usepackage{pgfplots}
\usepgfplotslibrary{statistics}
\usetikzlibrary{fadings}

\usepackage{xcolor,pifont}
\newcommand*\colourcheck[1]{%
	\expandafter\newcommand\csname #1check\endcsname{\textcolor{#1}{\ding{52}}\xspace}%
}
\newcommand*\colourcross[1]{%
	\expandafter\newcommand\csname #1cross\endcsname{\textcolor{#1}{\ding{56}}\xspace}%
}

\ifCLASSOPTIONcompsoc
\usepackage[caption=false, font=normalsize, labelfont=sf, textfont=sf]{subfig}
\else
\usepackage[caption=false, font=footnotesize]{subfig}
\fi

\definecolor{myyellow}{RGB}{255, 228, 26}
\definecolor{myblue}{RGB}{50, 50, 220}

\newboolean{showcomments}
\setboolean{showcomments}{true}
\ifthenelse{\boolean{showcomments}}
{
	\newcommand{\nb}[2]{
		{\sf
			\fcolorbox{myyellow}{yellow}{\scriptsize\textbf{#1}}%
			$\blacktriangleright$%
			{\color{myblue}\fontsize{7pt}{8pt}\selectfont\textbf{#2}}%
		}%
	}
}
{
	\newcommand{\nb}[2]{}
}

\newcommand{\tool}{\textsc{GenBo}\xspace} %
\newcommand{\baseline}{(1+1)-EA\xspace} %
\newtheorem{definition}{Definition}

\newcommand{\head}[1]{\noindent\textbf{\textit{#1.}}}

\fancypagestyle{mystyle}{
	\fancyhf{} %
	
	\fancyhead[C]{\scriptsize This article has been accepted for publication in IEEE Transactions on Software Engineering. This is the author's version which has not been fully edited and content may change prior to final publication. Citation information: DOI 10.1109/TSE.2024.3420816}
	\fancyhead[R]{\thepage}
	\fancyfoot[C]{\scriptsize \textcopyright\ 2024 IEEE. Personal use is permitted, but republication/redistribution requires IEEE permission. See https://www.ieee.org/publications/rights/index.html for more information.} %
}

\begin{document}
	
	\title{
		Boundary State Generation for Testing and Improvement of Autonomous Driving Systems
	}

	\author{Matteo~Biagiola, %
		and~Paolo~Tonella,~\IEEEmembership{Member,~IEEE Computer Society}%

		\IEEEcompsocitemizethanks{%
				\IEEEcompsocthanksitem Matteo Biagiola and Paolo Tonella are with the Universit\`{a} della Svizzera italiana, Lugano, Switzerland. Street: Via Buffi 13. Postcode: 6900. E-mail: \{matteo.biagiola, paolo.tonella\}@usi.ch.
			}
	}

	\markboth{IEEE TRANSACTIONS ON SOFTWARE ENGINEERING,~Vol.~XY, No.~X, XYZ~2020}%
	{Biagiola \MakeLowercase{\textit{et al.}}: Boundary State Generation for Testing and Improvement of Autonomous Driving Systems}
		
	\IEEEoverridecommandlockouts
	\IEEEpubid{\makebox[\columnwidth]{\textcopyright\ 2024 IEEE. Personal use is permitted, but republication/redistribution requires IEEE permission. See https://www.ieee.org/publications/rights/index.html for more information.}
	}
	
	\IEEEtitleabstractindextext{%
	\begin{abstract}
		
		Recent advances in Deep Neural Networks (DNNs) and sensor technologies are enabling autonomous driving systems (ADSs) with an ever-increasing level of autonomy. However, assessing their dependability  remains a critical concern.
		State-of-the-art ADS testing approaches modify the controllable attributes of a simulated driving environment until the ADS misbehaves. 
		In such approaches, environment instances in which the ADS is successful are discarded, despite the possibility that they could contain hidden driving conditions in which the ADS may misbehave.
		
		In this paper, we present \tool (GENerator of BOundary state pairs), a novel test generator for ADS testing. \tool mutates the driving conditions of the ego vehicle (position, velocity and orientation), collected in a failure-free environment instance, and efficiently generates challenging driving conditions at the behavior boundary (i.e., where the model starts to misbehave) in the same environment instance. 
		We use such boundary conditions to augment the initial training dataset and retrain the DNN model under test. Our evaluation results show that the retrained model has, on average, up to 3$\times$ higher success rate on a separate set of evaluation tracks with respect to the original DNN model.
		
	\end{abstract}
	
	\begin{IEEEkeywords}
		Software Testing, Autonomous Driving
\end{IEEEkeywords}}

	\maketitle
	
	\pagestyle{mystyle}
	\thispagestyle{mystyle}
	
	\IEEEdisplaynontitleabstractindextext

	\IEEEpeerreviewmaketitle
	
	\section{Introduction} \label{sec:introduction}

While the dream of fully-autonomous vehicles (the so-called \textit{Level 5}, as defined by the Society of Automotive Engineers (SAE)~\cite{SAE}) is still to come, there exist deployed systems that exhibit impressive levels of automation in the driving task (e.g., the Tesla's autopilot is considered an advanced Level 2 system~\cite{techspot}). Hence, testing of autonomous driving systems (ADSs) has become a critical research topic, of vital importance for today's and for future ADSs.

The literature on testing ADS systems is quite rich. A recent survey on ADS testing~\cite{survey-ads} analyzed 181 papers  published in peer-reviewed venues between June 2015 and June 2022. Most of the papers target vision-based ADSs that use Deep Neural Networks (DNNs) to process the input images and decide the actions to take. Such systems are typically trained in a supervised fashion using a dataset of labeled images, where the labels determine the desired actions (e.g., steering angle and throttle). DNN vision-based ADSs are tested offline and/or online~\cite{offline-fitash-emse,stocco-offline-vs-online}. Online testing usually involves a simulator (hence, it is often called simulation-based testing), where the DNN model drives the vehicle with the objective of keeping it within the driving lane~\cite{stocco-misbehaviour,asfault,deepjanus,deephyperion}.

Existing simulation-based testing approaches modify the driving environment in which the DNN model operates. A test case is a set of values representing all (or a subset of) the controllable attributes of the environment. Such attributes determine the environment  in which the ADS operates. In particular, a test case can be a road network~\cite{asfault}, a sequence of waypoints determining the track shape~\cite{deepjanus,deephyperion} or a mixture of dynamic and static elements of the simulation~\cite{morlot,samota,icse18-abdessalem,calo-generating}. Examples of static elements that do not change during the simulation, are the track shape, or the positions of the buildings and trees; dynamic elements are the weather condition and the dynamics, including position and velocity, of the other actors in the environment (e.g., pedestrians and other vehicles).

Current ADS testing approaches generate new environment scenarios until the ADS under test misbehaves, discarding those where the ADS is successful. Nonetheless, even in such \textit{failure-free} scenarios, there might exist hidden driving conditions that are left unexplored, and would potentially represent further opportunities to challenge the ADS. This calls for new ADS testing approaches designed to generate challenging driving conditions within the same environment instances. Such testing approaches would also be amenable in resource-constrained settings, where only a limited number of environment configurations (e.g., driving tracks) are available~\cite{waymo,borg,stocco-mind}, as opposed to existing test generators that modify the simulation environment. Indeed, changing the environment in such settings is expensive (e.g., building new testing tracks) and, sometimes, impossible (e.g., changing the weather condition).

\begin{figure*}[ht]
\centering

\includegraphics[trim=1cm 45cm 0cm 1.5cm, clip=true, scale=0.15]
{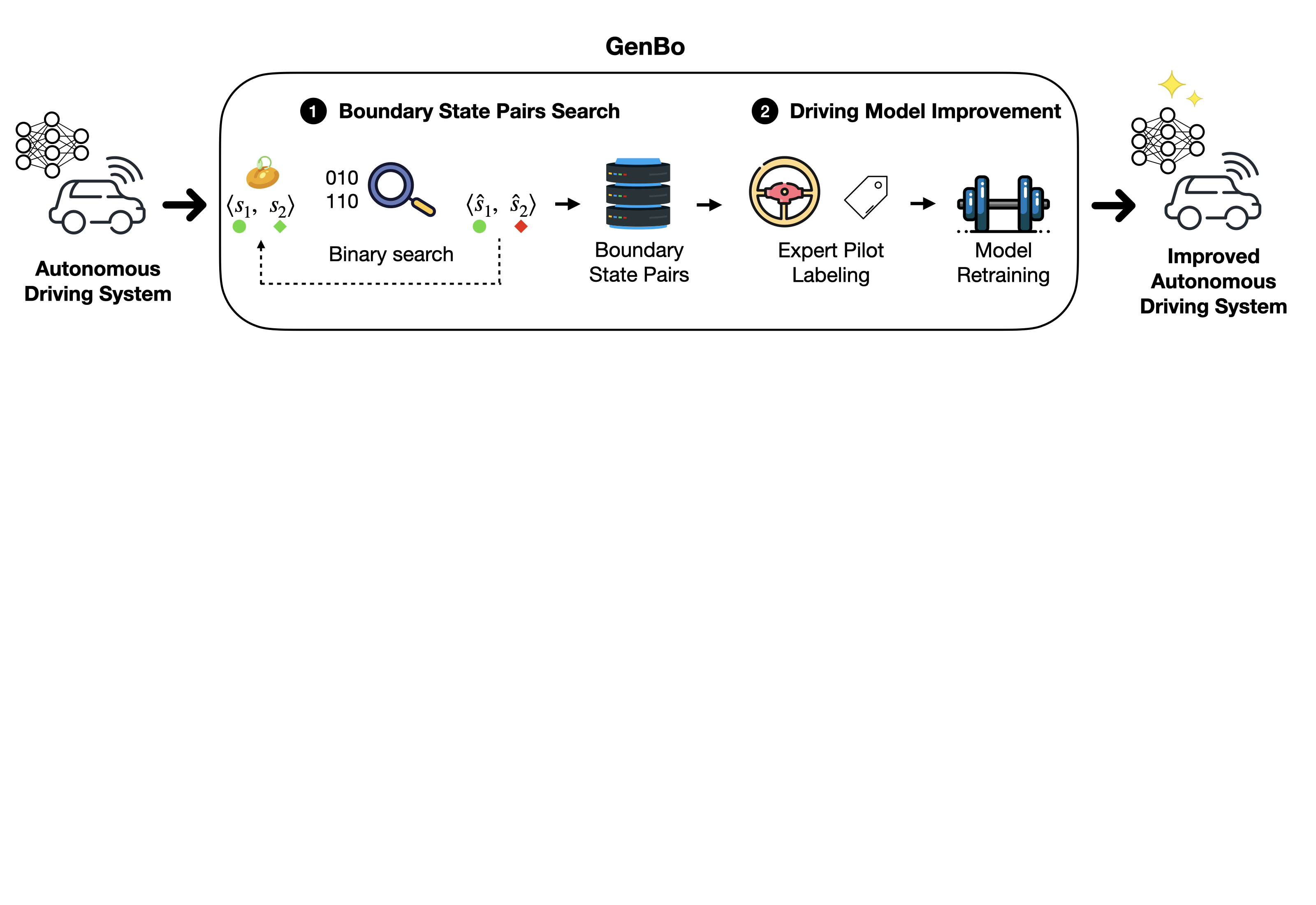}

\caption{Overview of our approach. Our approach takes an Autonomous Driving System (ADS) as input and features two steps, namely \ding{182}~\textbf{Boundary State Pairs Search} and \ding{183}~\textbf{Driving Model Improvement}. The first step~\ding{182} looks for boundary state pairs of the driving model controlling the ADS, while the second step~\ding{183} retrains the driving model using data labeled by an expert pilot driving on boundary state pairs. The output of the approach is an ADS improved by the boundary state pairs.} 
\label{fig:approach:overview} 
\end{figure*}

We propose \tool (\textbf{GEN}erator of \textbf{BO}undary state pairs), a novel test generator for online ADS testing. \tool works by mutating the driving conditions of the ego vehicle the DNN model is controlling, i.e., position, velocity and orientation (a \textit{state} hereafter), within a fixed environment scenario (including the track shape, the weather condition, etc.).
\tool makes full use of an existing, failure-free environment instance, by uncovering challenging driving conditions typically neglected by state-of-the-art approaches. In particular, \tool employs a novel search algorithm that evolves pairs of states in order to find \textit{boundary state pairs}, i.e., pairs of states that are \textit{close} to each other and that expose different (successful vs failing) behaviors of the ADS~\cite{deepjanus,mullins-frontier,tuncali-frontier}. \tool uses mutation operators to generate a sequence of state pairs that crosses the behavioral boundary and then applies binary search, which executes a logarithmic number of driving simulations, to efficiently reach the target boundary state pairs.

Among the boundary state pairs, we are interested in the \textit{recoverable} ones, i.e., those where an expert pilot can avoid the failure the ADS exhibits. \tool uses such pairs to collect a dataset of challenging driving conditions, labeled by an expert pilot.
Our hypothesis is that by augmenting the original training dataset with such challenging driving scenarios we can increase the generalization capabilities of the ADS to related, but unseen, driving tracks, which contain similar boundary conditions.

We applied our approach to test the lane-keeping functionality of the Dave-2 model~\cite{nvidia-dave2}, a test subject widely used in the related literature on ADS testing~\cite{third-eye,stocco-mind,stocco-jsep,stocco-misbehaviour,deephyperion,deepjanus,sbft-2023,deepguard,adept}. In our empirical evaluation, we tested driving models from different stages of training, and we show that \tool can find boundary state pairs even for well-trained models. Our results also show a very strong discriminative capability of boundary state pairs, as boundary state pairs of well-trained models are significantly more challenging than those of poorly trained models. Moreover, we retrained two high-quality driving models after augmenting the original training dataset with labeled examples collected from their boundary state pairs. Results show that the success rate of the retrained models on a set of evaluation tracks is on average up to 3$\times$ higher with respect to that of the original model.

Our paper makes the following contributions:

\begin{description}[noitemsep]
	\item [Technique.] A novel approach, implemented in the publicly available tool \tool~\cite{replication-package}, which exploits a limited number of failure-free environment instances to generate challenging driving conditions for the driving model under test.
	\item [Evaluation.] An empirical study showing that \tool exposes challenging driving conditions even for well-trained driving models and that such conditions are useful to significantly  improve the model.
\end{description}

	\section{Approach} \label{sec:approach}

\autoref{fig:approach:overview} shows an overview of our approach, which takes as input an ADS and performs two steps: step \ding{182} executes a search algorithm that extracts boundary state pairs from the driving model of the ADS, i.e., its decision-making component. In particular, the search algorithm evolves a pair of nearby states characterizing the driving conditions of the vehicle on a given environment scenario. In the initial pair of states $\langle s_1, s_2 \rangle$ (also called seed state), the driving model succeeds, i.e., when it starts to drive the vehicle either in state $s_1$ or $s_2$, it is able to successfully drive it along the driving track of the given scenario (i.e., none of the two starting states causes a failure). The objective of the search algorithm is to find a pair of \textit{boundary} states $\langle \hat{s}_1, \hat{s}_2 \rangle$, such that in one state (say $\hat{s}_1$) the driving model succeeds and in the other ($\hat{s}_2$) it fails. Once the search finds a boundary state pair, it samples a new seed until there is search budget, to find multiple boundary state pairs for the given driving model.

\begin{figure}[H]
\centering

\includegraphics[trim=0cm 51cm 87cm 0cm, clip=true, scale=0.23]
{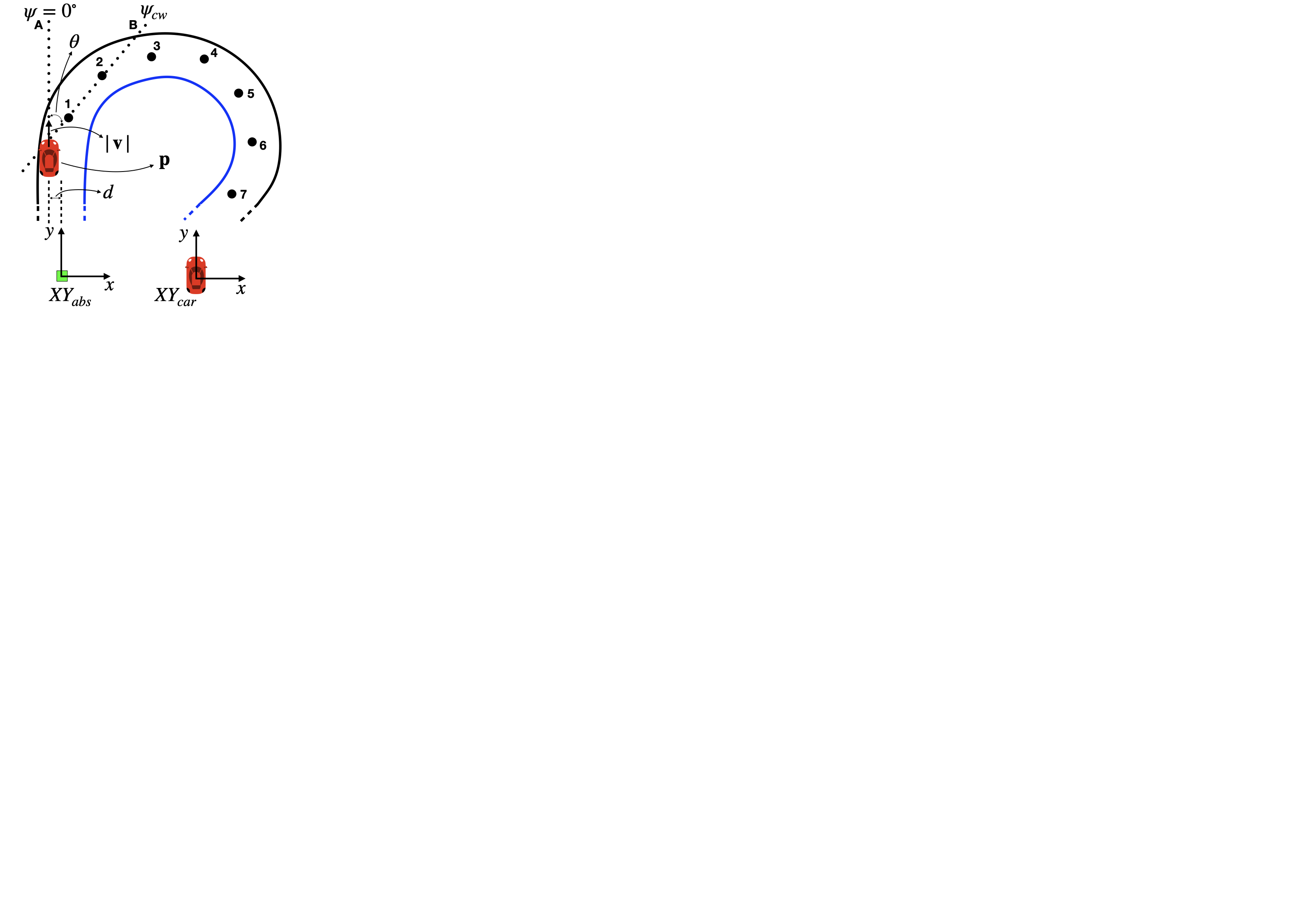}

\caption{Visualization of a vehicle state. A vehicle state is defined by three key properties, namely the position of the vehicle inside the driving track $\mathbf{p}$, the orientation of the vehicle $\psi$, and the magnitude of the velocity vector $v = | \mathbf{v} |$.} 
\label{fig:approach:state} 
\end{figure}

Boundary state pairs identify challenging driving conditions for the driving model in a given environment scenario, which can help to improve its capabilities in other scenarios where similar challenging conditions may occur. In the second step~\ding{183} of our approach, we resort to an expert pilot to drive on the boundary state pairs of the driving model.
The expert pilot labels the new camera images with the ground-truth steering angle values and outputs a labeled dataset. We use such dataset to retrain and improve the driving model of the ADS.

In our experiments, we instantiate the expert pilot with an autopilot that has global knowledge of the given environment scenario. The autopilot automates the labeling process in simulation, as it relies on precise information such as the position of the vehicle at each timestep, and the positions of the track waypoints where the vehicle drives. When the autopilot is not available, e.g., in the real world, the expert pilot is the human driver, labeling the camera images by driving on the road~\cite{nvidia-dave2, ieee-autopilot-data}.

\subsection{Boundary State}

The state of a vehicle driving along a track within an environment scenario (i.e., a set of variables representing the state of static and dynamic objects) is a subset of variables characterizing its motion. In particular, a test scenario in our approach is a vehicle state within a fixed environment  (in our evaluation  a closed driving track with an asphalt road surrounded by green grass, and sunny weather). We define the vehicle state as follows:

\begin{definition}[\textbf{State}] \label{definition:approach:state}
	A state $s$ of a vehicle in a track $T$ is a tuple $\langle \mathbf{p}, \psi, v \rangle$ where:
	\begin{enumerate}
		\item $\mathbf{p}$ is the $\langle x, y \rangle$ absolute position of the vehicle inside $T$ (w.r.t. the frame of reference $XY_{abs}$);
		\item $\psi$ is the orientation of the vehicle;
		\item $v = |\mathbf{v}|$ is the magnitude of the velocity vector $\mathbf{v}$ of the vehicle.
	\end{enumerate}
\end{definition}

\autoref{fig:approach:state} shows an example of a vehicle state. The black dots in the figure indicate the waypoints of the track, while the green square indicates the origin of the track. The waypoints are evenly spaced along the track, and they are placed at the center. There are two reference systems, i.e., the absolute reference system $XY_{abs}$ and the reference system of the vehicle $XY_{car}$. All the quantities that follow are defined w.r.t. $XY_{abs}$, except for the components of the velocity that are defined w.r.t. $XY_{car}$.  

The vehicle has position $\mathbf{p}$, where the vector $\mathbf{p}$ is centered at the center of the mass of the vehicle. It has a magnitude velocity of $v = |\mathbf{v}|$~\footnote{In practice, in a rigid-body simulator, setting the initial velocity of an object has an immediate effect~\cite{unity-velocity}; afterward, the velocity of the object is determined by the acceleration produced by the driving model controlling it.} and orientation $\psi$. The orientation is defined w.r.t. the $y$ axis of $XY_{abs}$ whose orientation is $0\degree$. Note that  acceleration cannot be included as part of the state since it is one of the variables controlled by the driving model. Consequently, at each step, the driving model determines the value of  acceleration, disregarding any initial acceleration value possibly provided with the state.

Given the state of a vehicle, we define a function $d(\mathbf{p}, T)$ that measures the distance $d$ of the vehicle from the center of the driving lane of track $T$ (also called cross-track error or XTE~\cite{stocco-mind}). It is also convenient to define a function $\theta(\mathbf{p}, \psi, T)$ that measures the angle $\theta$ between the orientation of the vehicle $\psi$ and the orientation of the closest waypoint $\psi_{cw}$ (the latter defining the direction of the road in a specific point in the track).
In the figure, the line crossing the vehicle is parallel to the $y$ axis of $XY_{abs}$, i.e., $\psi = 0\degree$ (see the dotted line ``A''). The next closest waypoint to the vehicle is waypoint ``1'' and its orientation is determined by the line going through the two successive waypoints (see the dotted line ``B''). The angle between the orientation of the vehicle and the orientation of the closest waypoint is $\theta$, the relative orientation of the vehicle.

\begin{definition}[\textbf{State Validity}] \label{definition:approach:state-validity}
	A state $s = \langle \mathbf{p}, \psi, v \rangle$ of a vehicle in a track $T$ is valid iff:
	\begin{enumerate}
		\item $d(\mathbf{p}, T) \leq \frac{W}{2}$, where $W$ is the lane width of track $T$;
		\item $v \leq v_{max}$;
		\item $\theta(\mathbf{p}, \psi, T) \leq \theta_{max}$;
	\end{enumerate}
\end{definition}

The first condition predicates that the center of the mass of the vehicle must stay within the bounding box of the track (i.e., the black and blue lines in \autoref{fig:approach:state}).
The second and third validity conditions regard velocity and relative orientation. In both cases, such quantities must be less than a maximum value, to be configured based on the mechanical properties of the autonomous vehicle being simulated. The three validity conditions ensure that a valid state is reachable by the driving model, as they prevent the search algorithm from exploring states that are trivially difficult for the driving model, given the physical properties of the vehicle it controls.

\begin{definition}[\textbf{Boundary State Pair}] \label{definition:approach:boundary}
	Given a driving model $M$ and a track $T$, a boundary state pair is a pair of valid states $\langle s_i, s_j \rangle$ such that:
	\begin{enumerate}
		\item the driving model $M$, when placed in state $s_i$, \textit{succeeds} at the task of driving the vehicle along track $T$;
		\item on the other hand, the driving model $M$, when placed in state $s_j$, \textit{fails} at the task of driving the vehicle along track $T$;
		\item the two states of the pair, i.e., $s_i$ and $s_j$, are distinct but close to each other, i.e., $s_i \neq s_j$ and \textit{dist}$(s_i, s_j) \leq \bm{\epsilon}$.
	\end{enumerate}
\end{definition}

This definition introduces the operation of \textit{placing} the driving model, and hence the vehicle, in a state $s$. For instance, if one of the two states in a boundary state pair is $s = \langle \langle 10, 0.5 \rangle,  0, 25 \rangle$, the simulator would place the vehicle in position $\mathbf{p} = \langle 10, 0.5 \rangle$, with an orientation of $\psi = 0 \degree$ w.r.t. $XY_{abs}$ and an initial velocity  $v = 25 \, km/h$. 

In each of the two states of a boundary state pair, we check if model $M$ is able to keep the vehicle in lane for a non-negligible amount of time $t \geq t_{min}$. For instance, if the driving simulation is set at 20 \textit{fps} and $t_{min} = 12.5 \, s$, \textit{succeeding} at the lane-keeping task means that the driving model $M$ is able to keep the vehicle in lane for at least $250$ simulation steps. In this case, we say that the state $s$ is a \textit{recoverable} state for the driving model $M$. 
On the other hand, \textit{failing} at the lane-keeping task, means that the vehicle goes out-of-bound before reaching the time threshold. In this case, we say that the state $s$ is a \textit{non-recoverable} state for the driving model $M$. 
A boundary state pair consists of one recoverable and one non-recoverable state.
In \autoref{fig:approach:overview}, we indicate recoverable states with a green circle/diamond and non-recoverable states with a red circle/diamond.

The last condition for two states to form a boundary state pair is \textit{closeness}:

\begin{equation}
	\textit{dist}(s_i, s_j) \leq \bm{\epsilon} \iff  \begin{cases} 
		\lVert \mathbf{p}_i - \mathbf{p}_j \rVert \leq \epsilon_p \quad  \\
		|v_i - v_j| \leq \epsilon_v \quad  \\
		|\psi_i - \psi_j|_{360 \degree} \leq \epsilon_\psi
	\end{cases}
\end{equation}

\noindent where $\bm{\epsilon} = \langle \epsilon_p, \epsilon_v, \epsilon_\psi \rangle$.
The first component of the distance  is the Euclidean distance between the two 2D position vectors of the two states. The second distance is the difference in absolute value between the magnitude velocity vectors of the two states. The third distance is the difference between two angles measured in degrees. To ensure that the difference stays within the interval $[0, 360]$, we take the modulo on $360 \degree$ (we use the notation $|\cdot|_{360\degree}$ to indicate this). The epsilon values, defining how close two boundary states should be, are configured depending on the properties of the track and of the autonomous vehicle being simulated.

\subsection{Boundary State Pairs Search}

\begin{algorithm}[t]
	
\DontPrintSemicolon
\footnotesize

\SetKwInOut{Input}{Input}
\SetKwInOut{Output}{Output}
\SetKw{KwBreak}{break}
\SetKw{KwContinue}{continue}
\SetKwRepeat{Repeat}{repeat}{until}

\Input{
	$M$, Driving model under test; \\
	\, $T$, Track to drive; \\
	\, $S$, Simulator instance; \\
	\, $Tr$, Reference trace; \\
 	\, $R$, Number of restarts; \\
	\, $N$, Number of iterations; \\
	\, $L$, Length of the sequence when mutating a seed state. \\
}
\Output{
	$A$, archive of boundary states.
}

$A$ $\gets$ $\emptyset$ \\
\For{$\textit{num\textsubscript{--}restarts} = 1$ \KwTo $R$} {
	\HiLi{/* Initialize seed state */} \\
	$s_1$ $\gets$ \textsc{sampleValidState}($Tr$) \\
	\Repeat{$\textit{valid}$} {
		$s_2$, $\textit{valid}$ $\gets$ $s_1$.\textsc{mutate}()
	}
	$b$ $\gets$ $\langle s_1, s_2 \rangle$ \\
	$\textit{success}_{1}$, $\textit{success}_{2}$ $\gets$ \textsc{execute}($b$, $S$, $T$, $M$) \\
	\If{ $\textit{success}_{1} \otimes \textit{success}_{2}$ $\land$ $b \notin A$}{ 
		$A$ $\gets$ $A$ $\cup$ $b$ \Comment*[r]{Store collateral boundary state}
	}
	\If{ $\neg \textit{success}_{1} \vee \neg \textit{success}_{2}$ }{ 
		\KwContinue \Comment*[r]{Restart if any execution failed}
	}
	
	\HiLi{/* Evolve seed state */} \\
	\textit{pairs} $\gets$ $[ \, b \, ]$ \\
	$\textit{num\textsubscript{--}iterations}$ $\gets$ $1$ \\
	\While{$\textit{num\textsubscript{--}iterations}$ $\leq$ $N$} {
		\For{$\textit{seq\textsubscript{--}length} = 1$ \KwTo $L$} {
			$\hat{b}$, \textit{valid} $\gets$ \textsc{getLast}(\textit{pairs}).\textsc{mutate}() \\
			\If{$\neg$ \textit{valid}}{
				\KwBreak
			}
			\textit{pairs}.\textsc{append}($\hat{b}$) \\
		}
		\textit{idx, it} $\gets$ \textsc{binarySearch}(\textit{pairs}, $S$, $T$, $M$) \\
		$\textit{num\textsubscript{--}iterations}$ $\gets$ $\textit{num\textsubscript{--}iterations}$ + \textit{it} \\
		\If{\textit{idx} $>$ $0$ $\land$ $\textit{pairs}[ \, \textit{idx} \, ]$ $\notin$ $A$}{
			$A$ $\gets$ $A$ $\cup$ $\textit{pairs}[ \, \textit{idx} \, ]$ \\
			\KwBreak
		}
	}
}
\Return{$A$}

\caption{Pseudocode of \tool}
\label{algorithm:approach:search}
\end{algorithm}

\autoref{algorithm:approach:search} shows the high level steps of the search algorithm. The algorithm evolves one individual at a time, where the individual is a pair of states. 

The main parameters of the algorithm are the number of iterations $N$, the number of restarts $R$ and the sequence length $L$. The number of restarts $R$ is the number of seed states that the algorithm samples and evolves (hence, $R$ is the maximum number of boundary states stored in the archive when the algorithm terminates). The number of iterations $N$ corresponds to the maximum number of executions of each individual for each restart. The sequence length $L$ is the maximum number of mutations of each individual in each iteration. This parameter needs to balance the mutation strength with the number of state pairs evaluation, avoiding overshooting the failure boundary while, at the same time, preventing the execution of too many state pairs that would unnecessarily consume the search budget. The output of the search algorithm is a set of boundary state pairs stored in the archive $A$.

The search algorithm consists of two main phases, i.e., the initialization of a seed state (lines 4--15) and its evolution (lines 17--33). The objective of the initialization phase is to generate a pair of states where the driving model succeeds; the subsequent evolution phase, starting from such pair of states, takes care of turning it into a boundary state pair. In the first phase, the algorithm starts by sampling a state from a reference trace $Tr$, by calling the \textsc{SampleValidState} function at line~4. The function also has a state validity filter, to ensure that the initial states are valid. A reference trace is a trajectory of states that we collect by running an autopilot with global knowledge of the track and of the state of the vehicle. The autopilot drives by following nominal trajectories i.e., it drives the vehicle at the center of the lane along the waypoints of the track. 
Alternatively, if the reference trace is not available, seed states can be sampled at random or can be obtained from a driving model $M$, with a downstream validity check that filters out invalid states.

The loop at lines~5--7 mutates the seed state $s_1$ until the resulting state $s_2$ is valid. As described below in \autoref{sec:state-mu}, the mutation operator ensures closeness of the boundary states by construction. At line~8, the algorithm builds the individual $b$ consisting of the two states $s_1$ and $s_2$~\footnote{We use the term ``individual'' to indicate a state pair, to follow the convention of the search-based literature where an individual is the object being evolved~\cite{mcminn-survey}.}. The \textsc{execute} function at line~9 places the vehicle in the two states of $b$, executes the two simulations, and collects the results. In particular, it returns two boolean values, $\textit{success}_1$ and $\textit{success}_2$, indicating whether the driving model $M$ succeeds at the lane-keeping task.
The \textbf{if} statement at lines~10--12, checks if the boolean values of $\textit{success}_1$, $\textit{success}_2$ have opposite values (XOR logical operator, indicated as $\otimes$), which indicates a boundary state pair has been found and stored into the archive $A$ (line~11) after checking for duplicates (i.e., $b \notin A$). When either of the two executions, or both, fail, we restart the search (lines~13--15), while the evolution loop (lines 17--33) is performed only when in both states of the pair $b$ the driving model succeeds.

The evolution phase starts by creating a list of pairs (line~17), initialized with the seed, and by assigning the counter variable for the following while loop (line~18). Such loop (lines~19--33) iteratively mutates the last pair in \textit{pairs} for $L$ times. The \textsc{mutate} function at line~21 outputs a new individual for every execution (i.e., $\hat{b}$), mutating both states. The function also returns a boolean variable (i.e., \textit{valid}), indicating whether it was possible to obtain a valid  state pair by mutation. If at any point during the sequence an invalid pair is generated, the algorithm breaks the \textbf{for} loop (\textbf{if} statement at lines~22-24) and goes directly to line~27. At this point the algorithm performs a binary search operation on the \textit{pairs} list to find a boundary state pair. The function \textsc{binarySearch} executes the last individual and if in both its states the driving model fails, it chooses an individual in the middle of the list. Then, it proceeds recursively with the head or the tail of the list depending on whether the middle individual consists of two states where the driving model fails or succeeds, respectively. When in one state the driving model fails and in the other it succeeds, a boundary state pair is found. The function \textsc{binarySearch} returns the number of executions of individuals into the variable \textit{it}. Moreover, it returns a positive index into the variable \textit{idx} if it found a boundary state pair. In such a case, the algorithm adds the corresponding boundary state pair ($\textit{pairs}[\, \textit{idx} \,]$) to the archive after checking for duplicates (\textbf{if} statement at lines~29--32) and restarts the search. If no boundary state is found, the \textbf{while} loop at lines~19--33 restarts by mutating the last individual of the \textit{pairs} list until the budget of $N$ iterations expires.

\subsubsection{State Mutation} \label{sec:state-mu}

In \autoref{algorithm:approach:search} there are two calls to the \textsc{mutate} function. The first one (line~6), mutates a single state into a new one, while the second one (line~21) mutates a pair of states. 
Both ensure closeness of the resulting state(s) to the initial one(s).

\begin{definition}[\textbf{State Mutation Function}] \label{definition:approach:state-mutation-function}
	Given two valid states $s_i, \, s_j \in \mathbb{S}$~\footnote{In case of the mutation function at line 6 in \autoref{algorithm:approach:search}, $s_j = s_i$.} that satisfy closeness (i.e., $\textit{dist}$($s_i$, $s_j$) $\leq$ $\bm{\epsilon}$), a mutation is a function $\mu: \mathbb{S} \times \mathbb{S} \rightarrow \mathbb{S}$ that modifies the state $s_i = \langle \mathbf{p}, \psi, v \rangle$ such that the resulting state $\hat{s}_i = \mu(s_i, s_j) = \langle \hat{\mathbf{p}}, \hat{\psi}, \hat{v} \rangle$ has the following properties:
	\begin{enumerate}
		\item $d(\hat{\mathbf{p}}, T) \geq d(\mathbf{p}, T)$;
		\item $\hat{v} \geq v$;
		\item $|\theta(\hat{\mathbf{p}}, \psi, T)| \geq |\theta(\mathbf{p}, \psi, T)|$;
		\item at least one of the three inequalities 1), 2), 3) is strict;
		\item $\textit{dist}$($\hat{s}_i$, $s_j$) $\leq$ $\bm{\epsilon}$;
		\item $\hat{s}_i$ is valid.
	\end{enumerate}
\end{definition}

The mutation function $\mu$ changes the position $\mathbf{p}$, the velocity $v$ or the orientation $\psi$ of the given state $s_i$, such that the resulting state has higher values, in absolute terms, of $d$, $v$ or $|\theta|$, while preserving closeness and validity. Changes to the velocity magnitude can be made directly on the given state, while in order to change the distance from the center $d$ and the relative orientation $\theta$, the mutation function $\mu$ needs to change the position $\mathbf{p}$ and/or the orientation $\psi$ of the vehicle. 

In the following, we describe the mutation function for each component of the state, i.e., position, orientation and velocity. After mutating one of these three components, chosen at random, a second and then a third mutation is applied to each of the two remaining components with probability 0.3. 

\head{Mutate position} The position vector $\mathbf{p}$ of the state $s_i$ being mutated has two components, i.e., $p_{x}$ and $p_{y}$. The position mutation operator changes either of the two components, chosen with equal probability (e.g., $p_{x}$). In order to respect the closeness condition, $\hat{p}_{x}$ is chosen in the interval $(p_{x} - \epsilon_p, p_{x} + \epsilon_p)$. Then, we determine the other component of $\hat{\mathbf{p}}$ (e.g., $\hat{p}_{y}$) by satisfying the following Euclidean distance inequality:

\begin{equation}
	\sqrt{(\hat{p}_{x} - q_{x})^2 + (\hat{p}_{y} - q_{y})^2} \leq \epsilon_p
\end{equation}

\noindent
where $q_{x}$ and $q_{y}$ are the $X$ and $Y$ components of the position vector of the other state in the boundary pair, $s_j$.
To make sure that the mutation increases the current XTE $d$, we randomly sample position components $\langle \hat{p}_{x}, \hat{p}_{y} \rangle$ from their respective ranges and compute the XTE $\hat{d}$ for each resulting position vector $\hat{\mathbf{p}}$. We use a finite budget to determine whether it is possible to obtain a higher XTE, while also checking that the  resulting state $\hat{s}_i$ is still valid. If the outcome of both checks is positive, we deem the mutation successful and change the position components of $s_i$ accordingly. 

\head{Mutate orientation} 
Given the orientation $\psi_i$ of state $s_i$, the orientation mutation operator applies a mutation such that the closeness and validity constraints hold.

\begin{figure}[h]
\centering

\includegraphics[trim=1cm 59cm 76cm 1cm, clip=true, scale=0.19]
{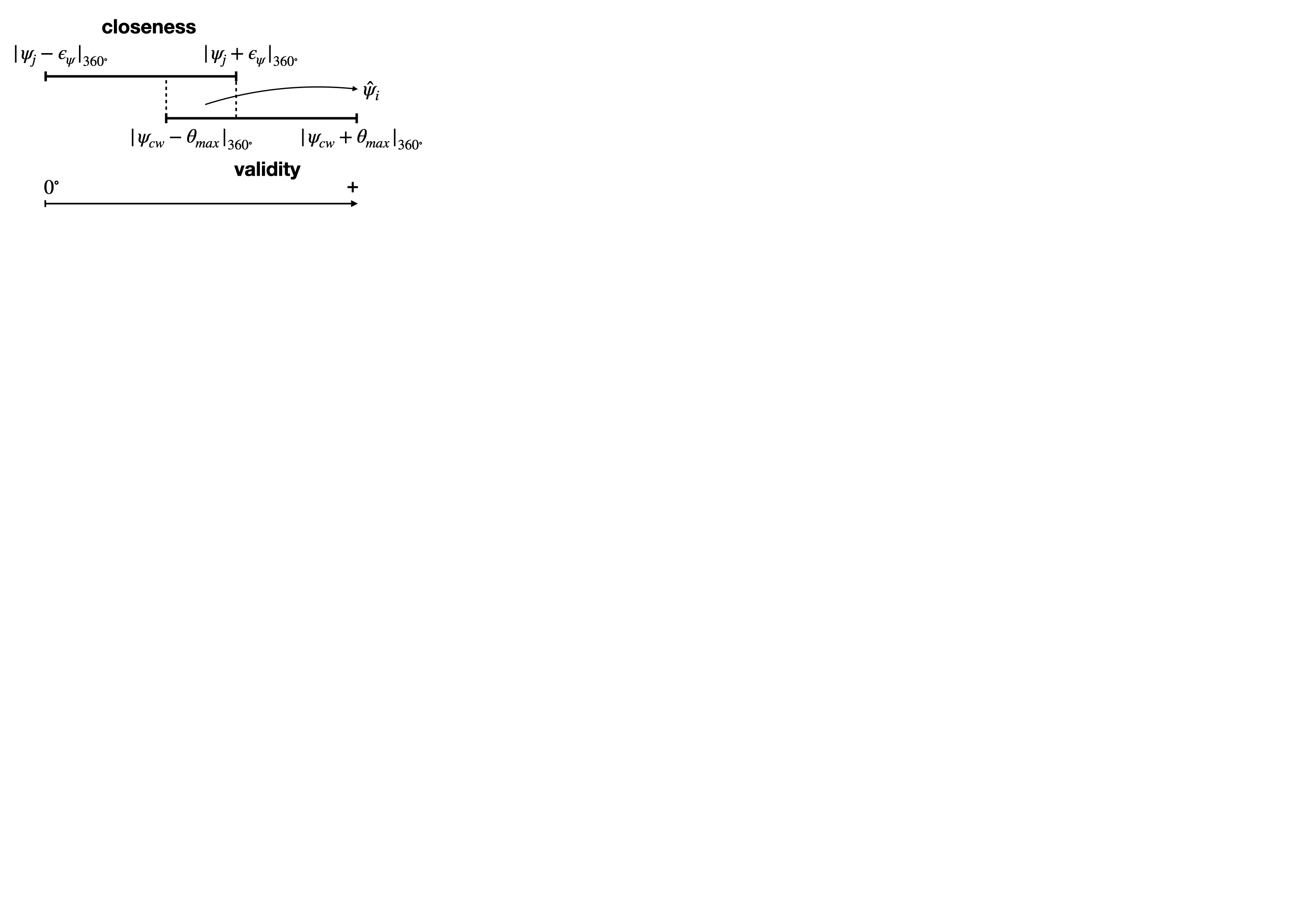}

\caption{Example of overlap between closeness and validity constraints concerning the ``mutate orientation'' function. Assuming that the orientation angle $\psi$ increases with the $x$ direction, the two ranges overlap, determining the new range for the mutated orientation $\hat{\psi}_i$ (delimited by the two dashed lines).} 
\label{fig:approach:orientation-overlap} 
\end{figure}

The closeness constraint regards the orientation of the other state of the pair, i.e., $\psi_j$, limiting the new orientation $\hat{\psi}_i$ to be $\epsilon_\psi$-close to $\psi_j$. On the other hand, the validity constraint on the mutated orientation $\hat{\psi}_i$ depends on the orientation of the closest waypoint $\psi_{cw}$, whose angle w.r.t. the new orientation must not be greater than $\theta_{max}$. In \autoref{fig:approach:orientation-overlap} there is overlap between the two ranges, hence we choose the new orientation $\hat{\psi}_i$ in the interval determined by the higher of the  two interval lower bounds (i.e., $|\psi_{cw} - \theta_{max}|_{360\degree}$ for the example in \autoref{fig:approach:orientation-overlap}) and the lower of the two  interval upper bounds (i.e., $|\psi_j + \epsilon_\psi|_{360\degree}$ for the example in \autoref{fig:approach:orientation-overlap}). 

In computing the intersection between closeness and validity intervals, we have to carefully consider that angles are defined modulo $360\degree$. For instance, with $\epsilon_\psi = 7.2$ (i.e., $360 \cdot 2\%$), $\theta_{max} = 20\degree$, $\psi_j = 350\degree$ and  $\psi_{cw} = 15\degree$, the resulting overlap range would be $[355\degree, 35\degree]$, and we would deem that there is no overlap as the upper bound is lower than the lower bound. To properly handle such cases, we have to represent the overlap range as $[355\degree, 360\degree] \cup [0\degree, 35\degree]$. Only by doing this we correctly find that in such case an overlap between validity and closeness ranges does exist, being the range $[355\degree, 357.2\degree]$.

When it is not possible to determine an overlap range, the mutation operator fails. We use a finite budget to randomly sample orientation values within the overlap range until the new relative orientation $\hat{\theta}$ is greater than the current relative orientation $\theta$. If the search finds such a value, we replace the orientation $\psi_i$ with new orientation $\hat{\psi}_i$ as it is respects all the constraints.

\head{Mutate velocity} The velocity mutation operator mutates the magnitude of the velocity vector. In particular the mutation needs to satisfy the following conjunction of constraints:

\begin{equation}
	|v_i - v_j| \leq \epsilon_v \land v_i \leq v_{max} \land \hat{v}_i > v_i
\end{equation}

\noindent
where the first constraint represents closeness, the second  validity, and the third  the increased velocity requirement. The constraint system is solvable if $v_i < v_{max}$; we randomly choose a solution  that satisfies all the constraints, $\hat{v}_i$, which is the mutated magnitude velocity. We then compute the components of the velocity vector $\mathbf{v}$ on the reference system of the vehicle $XY_{car}$. We set $v_x = 0, \, v_y = \hat{v}_i$, since only the vertical component of the velocity is non-zero in the coordinate system of the vehicle. To compute the absolute velocity vector on the reference system $XY_{abs}$ we carry out a rotation of axis using the orientation of the vehicle $\psi$~\cite{rotation-axes}. 

\begin{figure}[h]
\centering

\includegraphics[trim=0cm 48cm 74cm 0cm, clip=true, scale=0.18]
{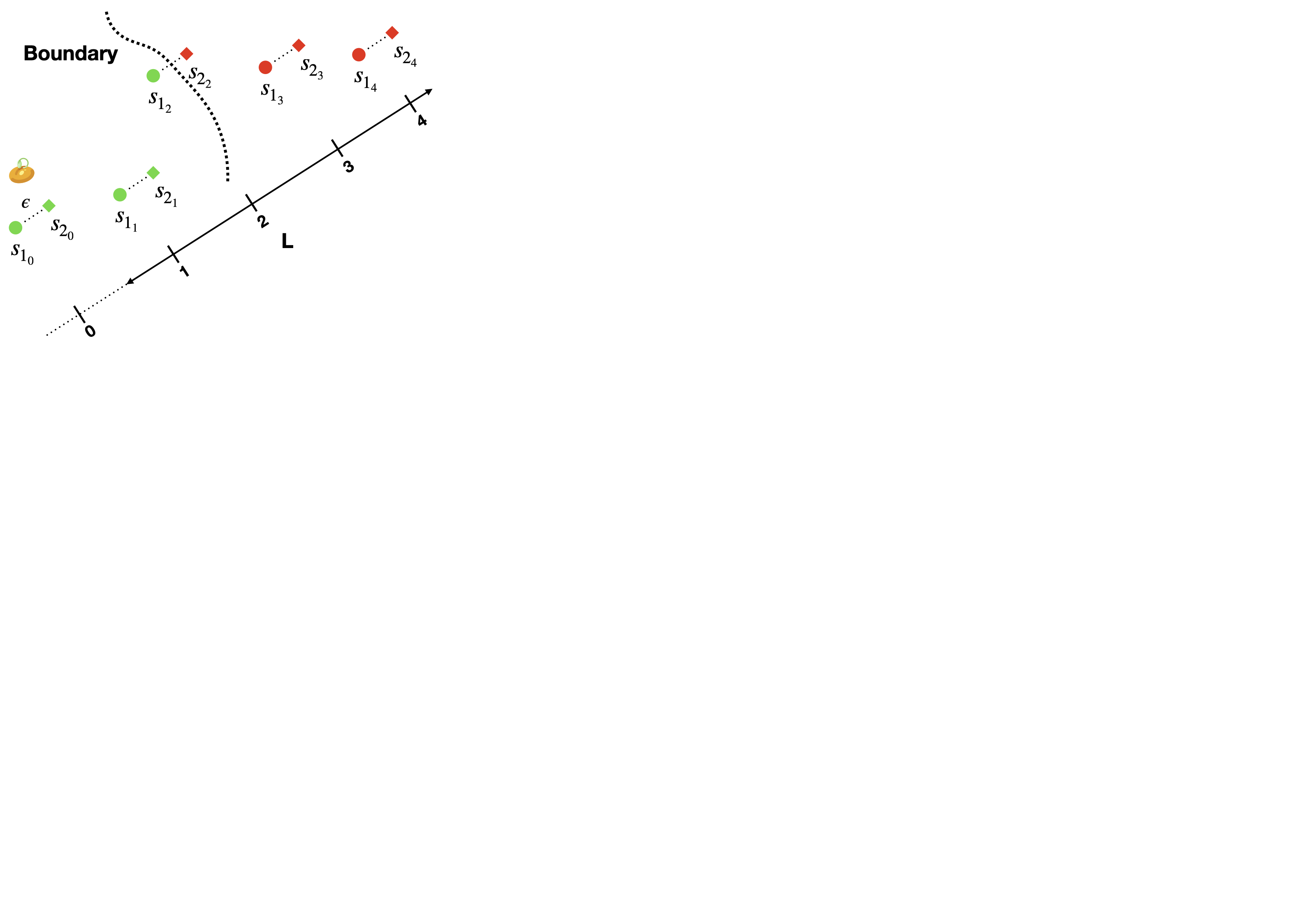}

\caption{Example of binary search to find a boundary state pair. The initial seed state is the pair $(s_{1_0}, s_{2_0})$, with $s_{2_0}$ being the more challenging state of the pair. The initial pair is consecutively mutated $L = 4$ times (difficulty increases going towards the right). Binary search efficiently finds the pair that sits on the boundary, i.e., $(s_{1_2}, s_{2_2})$}.
\label{fig:approach:binary-search} 
\end{figure}

\subsubsection{Binary Search}

In \autoref{algorithm:approach:search} at line~27, the search algorithm calls the binary search function on the sequence of mutations \textit{pairs} generated from the seed $b$. \autoref{fig:approach:binary-search} illustrates the process on a sequence of $L = 4$ mutations. Applying consecutive increasing mutations (i.e., mutations that increase either $d$ or $v$ or $\theta$) to a starting pair of states makes the pair increasingly more difficult for the driving model. We measure the difficulty of a state by using the maximum cross track error (XTE) across a given simulated scenario, which is defined as the distance between the center of mass of the vehicle and center of the lane~\cite{stocco-mind,maxibon}. The XTE is zero when the vehicle drives at the center of the lane and it is maximum when the vehicle goes out of bound~\footnote{We empirically validated the hypothesis of the state difficulty  increasing  monotonically with the mutation strength, by applying the Mann-Kendall trend test (with a significance level $\alpha = 0.05$)}~\cite{mann-nonparametric,kendall-rank} to the sequence of XTE values upon consecutive increasing mutations. Results show that there is an overall positive trend of XTE values at increasing mutation strength..

Depending on the length of the sequence, the resulting pairs will move away from the initial seed and potentially far from the boundary. The idea is to apply a binary search to look for the pair that sits on the boundary, given that the pairs in the sequence are sorted by their difficulty~\footnote{In the presence of non-monotonic behaviors, the algorithm returns a boundary state pair (if it exists)}, although it might not be the most challenging one..

In \autoref{fig:approach:binary-search}, the first state ($\textit{idx} = 0$) is the seed, where in both states the model $M$ is successful by construction. The last state ($\textit{idx} = 4$) is the most challenging  and the driving model fails in both states. Then, the binary search function executes the pair at index $\textit{idx} = 2$. In this example, the model succeeds in $\hat{s}_1$ and fails in $\hat{s}_2$. The two states in the pair are close and valid by construction, forming a boundary state pair. 

\subsubsection{Individual-level Mutation}

In the previous sections we described how to mutate a  state, but \autoref{algorithm:approach:search} evolves individuals that are state pairs, rather than single states. 
Given a state pair $b = \langle s_1, s_2 \rangle$,  the individual-level mutation function  starts by mutating $s_2$, producing a new valid state $\hat{s}_2$ close to $s_1$. Since the pair $\langle s_1, s_2 \rangle$ consists by construction of a less challenging state $s_1$ and a more challenging state $s_2$ (see \autoref{algorithm:approach:search}), during the evolution of an individual we want to preserve such a relation between the elements of the pair, which makes it easier to cross the failure boundary with the most challenging state ($\hat{s}_2$) while keeping the other state ($\hat{s}_1$) within the boundary (see \autoref{fig:approach:binary-search}).

Hence, we compute the difference operator $\delta_s = \hat{s}_2 - s_2 = \langle \bm{\delta}_p, \delta_{\psi}, \delta_{v} \rangle$ and  apply it to $s_1 = \langle \mathbf{p}_1, \psi_1, v_1 \rangle$, obtaining $\hat{s}_1 = \langle \mathbf{p}_1 + \bm{\delta}_p, \psi_1 + \delta_{\psi}, v_1 + \delta_{v} \rangle$. Then, we check whether $\hat{s}_1$ is valid. In particular, the position is the most critical parameter since it also affects the relative orientation. If $\hat{s}_1$ is not valid, e.g., because adding the deltas resulted in a position that is out of the lane or the orientation is invalid, the individual-level mutation operator discards both states $\hat{s}_2$ and $\hat{s}_1$, and re-attempts to mutate the pair until either both resulting states are valid or the budget expires.

\subsection{Driving Model Improvement}

The second step of our approach (i.e., step~\ding{183}) consists of improving the driving model of the ADS. We first place the expert pilot on the non-recoverable states of the boundary state pairs of the driving model. The expert pilot labels each frame of the simulation with the right steering angle, as it can avoid the failures that affect the ADS under test.

In the next phase, we retrain the driving model from scratch by using the union of the initial dataset used for training the model, and the labeled dataset resulting from the placing the expert pilot on the boundary state pairs of the driving model.

	\section{Empirical Evaluation} \label{sec:evaluation}

To assess the existence and the practical benefits of boundary state pairs, we consider the following research questions:

\noindent
\textbf{RQ\textsubscript{1} (Existence):} \textit{Do boundary state pairs exist in the training track for well-behaving driving models?}

In our analysis we are interested in evaluating \textit{well-behaving} driving models, i.e., models that, in nominal conditions, drive well in the driving scenario they have been trained on (in our evaluation, a closed driving track with asphalt roads, surrounded by green grass and sunny weather~\cite{deepjanus,deephyperion}). For nominal conditions, we mean that the vehicle is placed in the starting position of the driving track, with zero velocity and oriented in the direction of the road. We say that a driving model drives well in nominal conditions if it is able to keep the lane for 1200 simulation steps (which correspond to two consecutive laps and 1 minute of driving at 20 \textit{fps}).

RQ\textsubscript{1} aims at assessing whether such models exhibit boundary state pairs. A positive answer would imply that challenging driving scenarios exist even in a failure-free track where the driving model is well-behaving. Therefore, testing driving models does not necessarily require the manipulation of the environment (e.g., by generating new driving tracks). 

\head{Metrics} To assess the existence of boundary state pairs given a well-behaving driving model, we simply measure the number of boundary state pairs resulting from the search process.

\noindent
\textbf{RQ\textsubscript{2} (Comparison):} \textit{How do boundary state pairs of driving models of different qualities compare with each other? Do boundary state pairs discriminate between them?}

In RQ\textsubscript{2}, we compare boundary state pairs of driving models with different degrees of performance. All the driving models respect the requirement of being able to complete the training track in nominal conditions (i.e., they are well-behaving) but they differ in terms of the amount of time they have been trained (i.e., they have different validation losses). We aim to study the extent to which boundary state pairs discriminate between high-and low-quality driving models. We expect boundary state pairs of high-quality driving models to be more challenging and extreme than those of low-quality driving models.

\head{Metrics} We use two metrics for comparing boundary state pairs. Such metrics are system-level since the search process executes the driving model within the system (i.e., the vehicle driving along a track).

The first system-level metric we consider is \textit{recoverability}, that measures to what extent the driving model $M_A$ is able to recover from the boundary state pairs of the driving model $M_B$. For instance, let us suppose that the driving model $M_B$ has $N$ boundary state pairs in the training track. We measure the recovery percentage of the driving model $M_A$ in each of them, as the number of times $M_A$ succeeds when placed in a certain state, divided by the number of runs the model is executed in each state (to take into account the randomness of the simulation~\cite{simulation-survey-icst}). The recoverability of $M_A$ on the boundary state pairs of $M_B$ is the average recovery percentage across all $N$ boundary state pairs. We consider the recoverability of $M_A$ on both the recoverable and non-recoverable states of $M_B$. Hence, recoverability is a measure of how challenging the boundary state pairs of a certain driving model are for another driving model.

The second system-level metric we consider is the \textit{radius}. We define the radius at the state level. In particular, we consider the non-recoverable state of a boundary state pair, since it is more challenging. Given a state $s$ and the training track $T$ we consider three quantities: (1)~the distance from the center $d(\mathbf{p}, T)$, (2)~the velocity $v$, and (3)~the relative orientation $\theta(\mathbf{p}, \psi, T)$. Our hypothesis is that such quantities determine how difficult a state is. In order to compare boundary state pairs of different driving models we need to both normalize each quantity for each state and to consider a point of reference to compare with. We define the point of reference as $\mathbf{\Omega}: (d(\mathbf{p}, T) = 0 \, m,\, v = 0 \, km/h, \, \theta(\mathbf{p}, \psi, T) = 0\degree)$, i.e., the origin. We then normalize distance, velocity and relative orientation based on their maximum values, i.e., respectively $\frac{W}{2}$ for the distance, $v_{max}$ for the velocity, and $\theta_{max}$ for the relative orientation. Given a state $s$ and the vector of normalized quantities $\mathbf{q}$, we compute the radius as the distance between $\mathbf{q}$ and the reference (i.e., $||\mathbf{q} - \mathbf{\Omega}||$) and normalize by dividing by the maximum distance. If a driving model has $N$ boundary state pairs, the radius of the driving model is the average of all the radii for each boundary state pair. The closer the resulting radius is to 1, the better the driving model. The rationale is that a boundary state pair consists of two states that are close to each other and such that one of them is recoverable by the driving model. If the radius is high, then the boundary state pair is challenging, but nonetheless the driving model is able to recover in one of the states of the pair. 

We use the Mann-Whitney rank test~\cite{mann-whitney-u-test} to assess the statistical significance of the difference between the radii of different models and the Vargha-Delaney effect size (i.e., $\hat{A}_{12}$) to assess the magnitude of such difference~\cite{vargha-delaney}, as previous literature suggests~\cite{arcuri-hitchhiker}.

\noindent
\textbf{RQ\textsubscript{3} (Retraining):} \textit{How effective are boundary state pairs in improving the performance of a high-quality driving model?}

In RQ\textsubscript{3}, we investigate the usefulness of boundary state pairs for improving a well-behaving driving model. In particular, we consider the boundary state pairs of the best model according to both model-level (i.e., validation loss) and system-level metrics (i.e., recoverability and radius). Our hypothesis is that such boundary state pairs identify challenging driving scenarios that improve the generalization capabilities of the model on a separate set of evaluation scenarios. A positive answer to this question would imply that 
a fixed and failure-free environment scenario contains hidden states that are useful not only to test but also to improve the ADS.

\head{Metrics} To assess the usefulness of the boundary state pairs of a given model, we measure the success rates of the original model on a set of evaluation tracks different from the training one. We then compare them with the success rates of the model retrained with the boundary state pairs dataset on the same evaluation tracks. 
We use the Wilcoxon signed-rank test~\cite{wilcoxon} to assess the statistical significance of the difference between the success rates of the two models on the same evaluation tracks. As for the radius, we use the Vargha-Delaney effect size (i.e., $\hat{A}_{12}$) to assess the magnitude of the difference.

\subsection{Procedure} \label{sec:evaluation:procedure}

Our experimental procedure consists of: (1)~searching for boundary state pairs of different driving models (RQ\textsubscript{1} and RQ\textsubscript{2}) and computing recoverability and radius for the boundary state pairs of each of them (RQ\textsubscript{2}); (2)~improving the best driving model according to validation loss and recoverability/radius, by using its generated boundary state pairs (RQ\textsubscript{3}). 

Across all research questions, we compare our approach with a 1 + 1 evolutionary algorithm (which we indicate as \baseline, hereafter). Indeed, the state of the art in testing ADSs are search-based approaches~\cite{deepjanus,asfault,sbft-2023}; among such approaches we choose \baseline as it is the most efficient. Moreover, being based on a single individual, it is directly comparable to our approach that only evolves one state pair (i.e., one individual) at a time. In a given iteration, \baseline mutates the current individual; then, between current and mutated individuals, the individual selected for the next iteration is the one with the highest XTE. To make the comparison with our approach fair, the mutation operators of \baseline are the same as with our approach.

\subsubsection{Implementation and Test Object} We implemented our approach in a Python tool called \tool, which is publicly available~\cite{replication-package}. The test object of our study is the popular DNN-based model Dave-2~\cite{nvidia-dave2}. Dave-2 is a robust lane-keeping model that has been previously used in several testing works in the literature~\cite{third-eye,stocco-mind,stocco-jsep,stocco-misbehaviour,deephyperion,deepjanus,sbft-2023,deepguard,adept}. We use the Donkey Car\textsuperscript{\texttrademark} open-source framework~\cite{donkeycar,sdsandbox} built with Unity~\cite{unity}. The simulator has been used in previous work to test both supervised learning and reinforcement learning models~\cite{stocco-mind,testing-drl-agents}; it is also featured in a recent survey as one of the prevalent open-source simulation platforms for online ADS testing~\cite{survey-ads}.

\subsubsection{Model Training} The first step to train a DNN-based driving model is to collect a dataset of labeled images captured by the camera of the vehicle driving along a track. We used the default closed track provided by the simulator, and, as expert pilot, we resorted to an autopilot with global knowledge of the track to automatically label the images with steering angle values. We decide the throttle command at runtime, both for the DNN-based driving model and the autopilot, via a linear interpolation between the minimum velocity ($10 \, km/h$) and the maximum velocity ($30 \, km/h$) so that the vehicle decreases its velocity when the steering angle increases (e.g., in a curve)~\cite{maxibon}.

As autopilot, we used a PID controller~\cite{pid} and tuned the Proportional, Integral and Derivative constants for the specific track. We made the PID controller drive the vehicle for three laps of the track, collecting approximately 5\textit{k} labeled images. We used an 80/20 training/validation split to train the Dave-2 model until convergence (we stopped the training after 10 epochs of no improvements of the validation loss). Furthermore, we saved a checkpoint every epoch of validation loss improvement. Of those models we kept four models we used for testing, at decreasing  validation losses. The first model, i.e., $M_1$, is the first well-behaving model. The last model, i.e., $M_4$, is the model with the best validation loss. We selected the two models in between, i.e., $M_2$ and $M_3$, by looking at the validation loss. In particular, we kept a model whenever its validation loss decreased by at least 10\% w.r.t. the previously considered model (i.e., $M_1$ and $M_2$ respectively).

\begin{figure*}[ht]
\centering

\includegraphics[trim=7cm 51.5cm 7cm 1cm, clip=true, scale=0.18]
{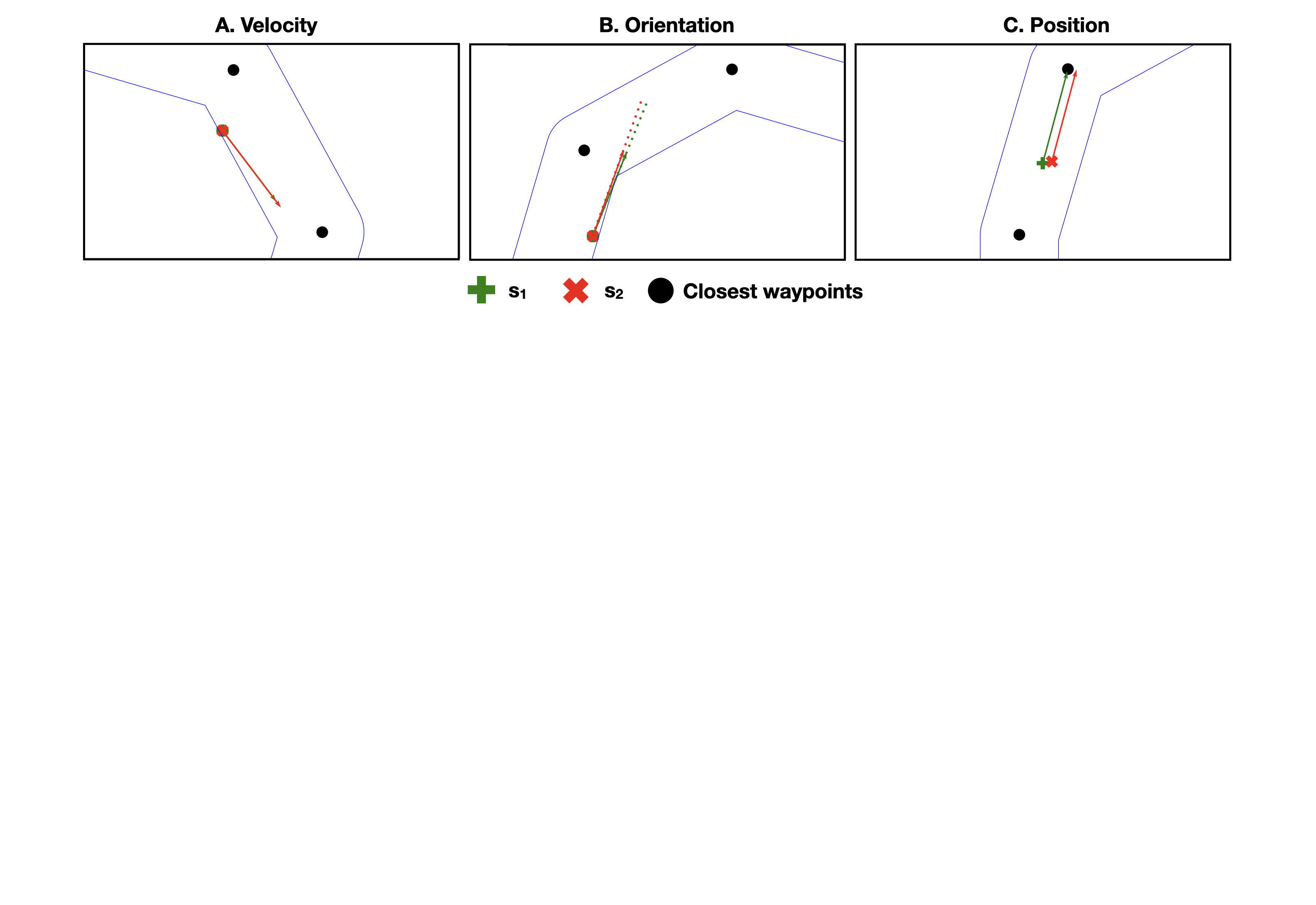}

\caption{Examples of boundary state pairs for the model $M_{\textit{last}}$. The two blue lines delimit the track $T$, the green plus sign is the state of the pair $s_1$ where the driving model succeeds, the red cross indicates the state $s_2$ where the driving model fails, and black dots are the closest waypoints (in terms of position within $T$) to the pair. The green and red arrows are the velocity vectors (their length is normalized and scaled 4$\times$ for visualization purposes).}
\label{fig:evalution:boundary-states-pairs} 
\end{figure*}

\subsubsection{Boundary State Pairs Search} We executed \tool and \baseline for the driving models obtained in the previous step, plus the autopilot, which we expect to perform better than the DNN-based models given its global knowledge of the track. We used the autopilot to collect the reference trace $Tr$ we use in \autoref{algorithm:approach:search} to sample valid states. As hyperparameters for \tool and \baseline, we used $R = 40$ number of restarts, $N = 10$ iterations for each restart, and we used $L = 3$ as sequence length for \tool. These choices were informed by preliminary experiments, where we found that they achieve a favorable trade-off between computation time and the effectiveness of the search in finding boundary state pairs for the most robust driving models. When placing the driving model on a certain state, we consider the state recoverable if the model is able to keep the vehicle in lane for at least 12.5$s$ which, at 20 \textit{fps}, corresponds to 250 simulation steps. We chose $\epsilon_p = W \cdot 10\%$, with $W = 4 \, m$ being the width of the lane in the track we use. We set the maximum velocity $v_{max} = 30 \, km/h$, and $\epsilon_v = v_{max} \cdot 10\%$. Regarding the orientation we chose $\epsilon_\psi = 360\degree \cdot 2\%$ and $\theta_{max} = 20\degree$. Such values of $\bm{\epsilon}$ ensure that the two states in the pair are \textit{reasonably} close to each other, effectively representing a boundary state pair if the driving model succeeds in one of the states and fails in the other.
To account for the randomness of the search algorithm~\cite{arcuri-hitchhiker} and of the driving simulation~\cite{simulation-survey-icst}, we executed \tool and \baseline nine times for each driving model. 

To specifically target the randomness of the simulation during the search~\cite{flakiness-driving-simulators,survey-robotics-icst}\footnote{Across different executions, the ADS might receive a slightly different sequence of images due to synchronization and timing inconsistencies.}, when the search algorithm finds a boundary state pair, we re-execute the driving model on each state of the pair three times. If the state pair is a boundary state pair the majority of the times (i.e., 2 out of 3 times), then we consider the state pair a boundary state pair and store it in the archive. At the end of the search process, we iterate over the boundary state pairs in the archive, and we compute a replication percentage across 10 repetitions, i.e., the number of times a state pair is actually a boundary state pair for the driving model under test out of 10 repetitions. For the subsequent analysis we consider as likely boundary state pairs those which have a replication percentage greater than zero. In fact, a boundary state pair is inherently flaky as the two states of the pair are \textit{close}, hence multiple repetitions are necessary to gather evidence of a boundary state pair status, i.e., non-zero boundary state replication percentage.

Each run of the search algorithm consists of at most $N \times R = 400$ driving simulations, plus a variable number of simulations due to replications depending on the number of boundary state pairs in the archive. If we were to discard the replications, we would have 400 $\times$ 5 driving models $\times$ 9 repetitions, consisting of a total of $\approx$18\textit{k} driving simulations per search algorithm (i.e., $\approx$36\textit{k} for \tool and \baseline).

In order to measure recoverability, we executed each driving model on the boundary state pairs of the others (both recoverable and non-recoverable), keeping track of the number of times a given model is able to recover from a certain state. Also in this case we consider a state recoverable, if the driving model is able to keep the vehicle in lane for at least 250 simulation steps.

\subsubsection{Driving Model Improvement} 
\autoref{table:evaluation:rq3} shows the list of tracks along with their features (Track Features macro-column).
In particular, we measured the curvature of the track (Column~1), i.e., the inverse of the minimum radius of the circles going through each sequence of three consecutive waypoints~\cite{deephyperion}, and the number of turns (Column~2), where a turn is a change of direction between consecutive waypoints by more than 5$^\circ$. The first track, i.e., $T_1$, is the training track, with a curvature of $0.29$ and 4 turns; the second track, i.e., $T_2$ is the specular version of $T_1$. To generate the other evaluation tracks, i.e., $T_3$---$T_9$, we randomly mutated the waypoints of $T_1$ until one of the following condition is true: (1)~the distance between $T_1$ and the candidate track $T_c$  (i.e., the Euclidean distance between the waypoints of the two tracks being compared) must be lower than a threshold; (2)~the curvature of $T_c$ must be greater than the curvature of $T_1$; (3)~the number of turns of $T_c$ must be greater than the number of turns of $T_1$. The distance constraint imposes that the resulting track is similar to $T_1$, while the curvature and the number of turns constraints ensure that it is more challenging to drive.

We executed the two best driving models (as they can no longer be improved by the training dataset as opposed to lower-quality models) on the evaluation tracks, $T_2$---$T_9$, to collect the success rates. We used 100 simulations, and we considered a simulation a success if the driving model is able to keep the vehicle in lane for 600 simulation steps (which approximately corresponds to one lap of the track at the maximum possible speed, and 30$s$ of driving at 20 \textit{fps}). For each search repetition of a search algorithm on a given driving model, we took the corresponding boundary state pairs in the archive and placed the autopilot in the non-recoverable state of each pair, in order to collect a labeled dataset for retraining. We executed the autopilot for 100 simulation steps per state and discarded the cases where the autopilot was not able to keep the vehicle in lane for that amount of time. Then, we merged the original training dataset and the labeled boundary state pairs dataset for each repetition, splitting the resulting dataset in such a way that the validation set is a superset of the validation set used for the initial training, to mitigate the possibility of regressions on the training track. We carried out retraining with the same hyperparameters of the initial training (random seed included) until convergence, obtaining, for a given driving model, nine retrained driving models, i.e., one for each search repetition (assuming that at least one boundary state pair was found by the search during that repetition). Finally, we executed each of them on the eight evaluation tracks to collect the success rates.

\subsection{Results} \label{sec:evaluation:results}

\begin{table*}[ht]
	\centering
	\begin{threeparttable}[b]
		
		\caption{Results for RQ\textsubscript{1} (Existence) and RQ\textsubscript{2} (Comparison). Bold-faced values indicate a statistically significant difference w.r.t. the radius of $M_1$; underlined values indicate a large $\hat{A}_{12}$. ``NA'' indicates that there are no boundary state pairs. Columns~2--3 show the boundary state pairs found by \tool and \baseline, while Columns~4--5 show the radius of such points. Columns~6--16 show the recoverability metric for all driving models on the boundary state pairs of \tool. Likewise, Columns~17--27 show the recoverability metric for \baseline's boundary state pairs. All values in the table are average across 9 repetitions.}
		\label{table:evaluation:rq1-rq2}
		\setlength{\tabcolsep}{3pt}
		\renewcommand{\arraystretch}{1.2}
			
			\begin{tabular}{r@{\hskip 1em}cccccccccccccccccccccccccc}
				\toprule
				
				\multicolumn{1}{l}{} & \multicolumn{2}{c}{\textbf{RQ\textsubscript{1}}} & \multicolumn{24}{c}{\textbf{RQ\textsubscript{2}}} \\
				
				\multicolumn{1}{l}{} & \multicolumn{2}{c}{\textbf{(Existence)}} & \multicolumn{24}{c}{\textbf{(Comparison)}} \\
				
				\cmidrule(r){2-3}
				\cmidrule(r){4-27}
				
				\multicolumn{1}{l}{} & \multicolumn{2}{c}{\# Boundary} & \multicolumn{2}{c}{\multirow{2}{*}{\rot{Radius}}} & \multicolumn{22}{c}{Recoverability (\%)} \\
				
				\cmidrule(r){6-27}
				
				\multicolumn{1}{l}{} & \multicolumn{2}{c}{States} & \multicolumn{2}{c}{} & \multicolumn{11}{c}{\tool} & \multicolumn{11}{c}{\baseline} \\
				
				\cmidrule(r){2-3}
				\cmidrule(r){4-5}
				\cmidrule(r){6-16}
				\cmidrule(r){17-27}
				
				\multicolumn{1}{l}{} & 
				\rot{\tool} & 
				\rot{\baseline} & 
				\rot{\tool} & 
				\rot{\baseline}& 
				\rot{$M_{1_{R}}$} &
				\rot{$M_{1_{NR}}$} & 
				\rot{$M_{2_{R}}$} & 
				\rot{$M_{2_{NR}}$} & 
				\rot{$M_{3_{R}}$} & 
				\rot{$M_{3_{NR}}$} & 
				\rot{$M_{\textit{last}_{R}}$} & 
				\rot{$M_{\textit{last}_{NR}}$} & 
				\rot{autopilot$_R$} & 
				\rot{autopilot$_{NR}$} & 
				\rot{Avg} & 
				\rot{$M_{1_{R}}$} & 
				\rot{$M_{1_{NR}}$} & 
				\rot{$M_{2_{R}}$} & 
				\rot{$M_{2_{NR}}$} & 
				\rot{$M_{3_{R}}$} & 
				\rot{$M_{3_{NR}}$} & 
				\rot{$M_{\textit{last}_{R}}$} & 
				\rot{$M_{\textit{last}_{NR}}$} & 
				\rot{autopilot$_R$} & 
				\rot{autopilot$_{NR}$} & 
				\rot{Avg} \\
				
				\midrule
				
				$M_1$ & 10.7 & 5.22 & 0.63 & 0.61 & -- & -- & 30 & 20 & 34 & 26 & 17 & 5 & 0 & 0 & 16 & -- & -- & 48 & 7 & 13 & 7 & 0 & 0 & NA & NA & 13 \\
				$M_2$ & 7.78 & 2.11 & \textbf{\underline{0.72}} & \textbf{\underline{0.65}} & 99 & 96 & -- & -- & 90 & 48 & 65 & 36 & 4 & 0 & 55 & 100 & 100 & -- & -- & 80 & 27 & 71 & 29 & NA & NA & 68 \\
				$M_3$ & 4.78 & 1.00 & \textbf{\underline{0.75}} & \textbf{\underline{0.69}} & 100 & 99 & 98 & 84 & -- & -- & 94 & 54 & 29 & 18 & 72 & 100 & 100 & 100 & 92 & -- & -- & 100 & 57 & NA & NA & 91 \\
				$M_{\textit{last}}$ & 5.89 & 0.78 & \textbf{\underline{0.73}} & 0.62 & 100 & 98 & 94 & 75 & 97 & 62 & -- & -- & 32 & 4 & 70 & 100 & 100 & 100 & 100 & 100 & 93 & -- & -- & NA & NA & 99 \\
				autopilot & 1.33 & 0.00 & \textbf{\underline{0.75}} & NA & 100 & 100 & 100 & 100 & 100 & 99 & 100 & 89 & -- & -- & 99 & 100 & 100 & 100 & 100 & 100 & 100 & 100 & 100 & -- & -- & 100 \\
				
				\midrule
				
				\textit{Avg} & 6.11 & 1.82 & 0.72 & 0.64 & 100 & 98 & 81 & 70 & 80 & 58 & 69 & 46 & 16 & 5 & -- & 100 & 100 & 87 & 75 & 73 & 57 & 68 & 46 & NA & NA & -- \\
				
				\bottomrule
				
			\end{tabular}
		
		\end{threeparttable}
\end{table*}

\subsubsection{Existence (RQ\textsubscript{1})}

\autoref{table:evaluation:rq1-rq2}, macro-column RQ\textsubscript{1} (Existence), shows the number of boundary state pairs the search process found for each driving model, averaged across the 9 repetitions. Both for \tool (Column~2) and \baseline (Column~3), we notice that the number of boundary state pairs decreases with the model quality, as measured by the validation loss. The only exception occurs for \tool between $M_3$ and $M_\textit{last}$, but with a small difference in terms of number of boundary state pairs found (the difference between the two is not statistically significant). 
As expected, it is challenging for the search algorithm to find boundary state pairs for the autopilot (on average 1.33 for \tool, while \baseline does not find any in all repetitions), since it is, by construction, robust against changes of the initial conditions of the vehicle. \tool finds on average six boundary state pairs across all driving models, against two of the \baseline approach, showing its higher effectiveness in extracting boundary state pairs, especially for the strongest driving models.

\autoref{fig:evalution:boundary-states-pairs} shows three examples of boundary state pairs for the best driving model $M_{\textit{last}}$. \autoref{fig:evalution:boundary-states-pairs}.A shows a \textit{velocity} boundary state pair, namely two states that have the same position and orientation, but with different velocity magnitudes (respectively $\approx$ 24 \textit{km/h} for $s_1$ and $\approx$ 27 \textit{km/h} for $s_2$); in this case, a slight change in velocity is enough for the driving model to lose control of the vehicle, resulting in an out-of-bound event. Similarly, \autoref{fig:evalution:boundary-states-pairs}.B shows an \textit{orientation} boundary state pair, where the two states differ only for their value of $\psi$ (respectively $21^\circ$ for $s_1$ and $19^\circ$ $s_2$); in this case, reducing the orientation towards the center of the track limits the agent's space and time to maneuver the vehicle while making the right turn. In \autoref{fig:evalution:boundary-states-pairs}.C, the position of the two states is different; in $s_2$ the vehicle is slightly ahead (both in the $x$ and $y$ directions) of $s_1$, and since the velocity in the two states is $\approx$ 30 \textit{km/h}, the agent is not able to make the turn in time.

\begin{tcolorbox}[colback=gray!15!white,colframe=black]
	\textbf{RQ\textsubscript{1} (Existence)}: Boundary state pairs exist for driving models with different performance. In particular, the number of pairs found by the search decreases with the quality of the driving model.
\end{tcolorbox}

\subsubsection{Comparison (RQ\textsubscript{2})}

\autoref{table:evaluation:rq1-rq2}, macro-column RQ\textsubscript{2} (Comparison), shows the two system-level metrics, i.e., \textit{Radius} (Columns~4--5) and \textit{Recoverability} (Columns~6--27), we used to compare the different driving models under test. All the values in the table (except the Avg columns and row), are averages across the 9 repetitions. Regarding the radius, we observe, for both \tool and \baseline, that the metric reaches a plateau after $M_2$, suggesting that it is able to discriminate low-quality driving models (i.e., $M_1$) from good to high-quality driving models (i.e., $M_2$---$M_\textit{last}$). For \tool the radius of the autopilot is in line with the DNN-based driving models $M_2$---$M_\textit{last}$ (there is no statistical difference between the radii of such driving models) while it is statistically different with a large effect size w.r.t. the radius of $M_1$ (as is the radius of the other driving models). The trend is similar for \baseline, where there is a statistically significant difference between the radius of $M_1$ and the radii of both $M_2$ and $M_3$, while, as opposed to \tool, there is no statistical difference between the radius of $M_1$ and the radius of $M_{\textit{last}}$. This is likely due to the fact that the radius of $M_{\textit{last}}$ is estimated with very few boundary state pairs (i.e., less than one per repetition on average), making the estimate less reliable. Overall, the radius seems to be sensitive to the number of boundary state pairs, especially for higher quality models (i.e., $M_2$, $M_3$, and $M_{\textit{last}}$), suggesting that more boundary state pairs adjust the radius estimate towards higher values.

Regarding recoverability, Column~6--15 shows the average recovery percentage for each driving model on the boundary state pairs found by \tool for the remaining models across the 9 repetitions. For instance, model $M_1$ has an average recovery percentage of 30\% on the recoverable states of $M_2$ (Column~8, $M_{2_{R}}$), while it has an average recovery percentage of 20\% on the non-recoverable states of $M_2$ (Column~9, $M_{2_{NR}}$). Looking at the \textit{Avg} row, we observe that the average recovery percentage decreases when the model quality increases. Indeed, the recovery percentage ranges from 100\% on the recoverable states of $M_1$ (i.e., $M_{1_{R}}$) to 46\% on the non-recoverable states of $M_\textit{last}$ (i.e., $M_{\textit{last}_{NR}}$). The recovery percentage further decreases when we consider the autopilot driving model. Its boundary state pairs are challenging for the DNN-based driving models, especially the non-recoverable states of the pairs, as the DNN-based driving models only recover on 5\% of them on average. The recoverability metric has a similar behavior for boundary state pairs found by \baseline. Indeed, the \textit{Avg} row shows a decreasing trend starting from 100\% recoverability on the pairs of $M_{1_R}$ to 46\% recoverability on the pairs of $M_{\textit{last}_{NR}}$

Column~16 shows the average recovery percentage of each driving model on the boundary state pairs found by \tool of the remaining models. We observe that the recovery percentage increases from 16\% of $M_1$ to 99\% of the autopilot. The recovery percentage of $M_2$ is more than twice as much that of $M_1$ (i.e., 55\%). The recovery percentage further increases when moving from $M_2$ to $M_3$ and then plateaus (the difference between $M_3$ and $M_{\textit{last}}$ is negligible, i.e., 2\%).
The same holds for the boundary state pairs found by \baseline (see Column~27). However, the average recoverability seems to be less granular than that measured by the boundary state pairs of \tool. In particular, the jumps between models of different qualities seem to be more pronounced (e.g., from 13\% to 68\% from $M_1$ to $M_2$ vs 16\%--55\% with \tool); moreover, the average recoverabilities of $M_{\textit{last}}$ and the autopilot are similar (i.e., respectively 99\% and 100\%), while the boundary state pairs of \tool better discriminate the two driving models (i.e., the average recoverability is respectively 70\% and 99\%). This difference might be due to the different number of boundary state pairs found by the two search strategies, which are then used to compute the recoverability metric; a higher number, as in the case of \tool, seems to suggest a more fine-grained computation of recoverability.

It is interesting to notice how our novel recoverability metric behaves when assessing the relative quality of two models. For instance, let us compare $M_1$ with $M_2$, when considering \tool as the search strategy (a similar comparison holds when considering \baseline): $M_1$ has a recoverability of 30\% (resp. 20\%) on the recoverable (resp. non-recoverable) states of $M_2$, while $M_2$ has a recoverability of 99\% (resp. 96\%) on the recoverable (resp. non-recoverable) states of $M_1$. This indicates a clear superiority of $M_2$ over $M_1$. The same pairwise comparison can be conducted for all pairs of driving models, consistently showing a very strong discriminative capability of the recoverability metric.

\begin{tcolorbox}[colback=gray!15!white,colframe=black]
	\textbf{RQ\textsubscript{2} (Comparison)}: Boundary state pairs discriminate driving models with different qualities. In particular, the radius discriminates low-quality from high-quality models at a gross granularity, while the recoverability metric is more fine-grained and discriminates very accurately the driving quality of each pair of models considered in our study.
\end{tcolorbox}

\subsubsection{Retraining (RQ\textsubscript{3})}

\begin{table}[ht]
	\centering
	\begin{threeparttable}[b]
		
		\caption{Results for RQ\textsubscript{3} (Retraining). Best success rates are highlighted in bold. Columns~2--4 show the features of the 9 evaluation tracks, such as curvature, number of turns and distance w.r.t. the original training track $T_1$. Columns~5--7 show the success rates for the $M_3$ driving model before retraining (Column~5) and after retraining, respectively with the boundary state pairs found by \tool (Column~6) and \baseline (Column~7). Likewise, Columns~8--10 show the success rates for the $M_\textit{last}$ driving model before and after retraining. The success rate values are averaged across 9 repetitions.}
		\label{table:evaluation:rq3}
		\setlength{\tabcolsep}{3pt}
		\renewcommand{\arraystretch}{1.2}
		
		\begin{tabular}{r@{\hskip 1em}ccc@{\hskip 1em}cccccc}
			\toprule
			
			\multicolumn{1}{l}{} & \multicolumn{3}{c}{Track Features} & \multicolumn{3}{c}{Success Rate (\%)} & \multicolumn{3}{c}{Success Rate (\%)} \\
			
			\cmidrule(r){2-4}
			\cmidrule(r){5-7}
			\cmidrule(r){8-10}
			
			\multicolumn{1}{l}{} & \multicolumn{1}{l}{} & \multicolumn{1}{l}{} & \multicolumn{1}{l}{} & \multicolumn{1}{l}{} & \rot{\tool} & \rot{\baseline} & \multicolumn{1}{l}{} & \rot{\tool} & \rot{\baseline} \\
			
			\multicolumn{1}{l}{} & Curv. & \# Ts. & Dist. & $M_3$ & $M^3_{\textit{retr.}}$ & $M^3_{\textit{retr.}}$ & $M_{\textit{last}}$ & $M^{\textit{last}}_{\textit{retr.}}$ & $M^{\textit{last}}_{\textit{retr.}}$ \\
			
			\midrule
			
			$T_1$\tnote{$\ast$} & 0.29 & 4 & -- & 100.0 & 100.0 & 100.0 & 100.0 & 100.00 & 100.0 \\
			$T_2$\tnote{$\dagger$} & 0.29 & 4 & -- & 16.00 & \textbf{75.00} & 56.00 & 0.000 & 0.000 & \textbf{41.43} \\
			$T_3$ & 0.34 & 5 & 20.13 & 0.000 & 0.000 & 0.000 & 56.00 & \textbf{72.00} & 0.143 \\
			$T_4$ & 0.32 & 5 & 19.08 & 0.000 & 0.000 & 0.000 & 0.000 & 0.000 & \textbf{11.57} \\
			$T_5$ & 0.28 & 5 & 29.31 & 21.00 & \textbf{86.56} & 70.60 & 70.00 & \textbf{93.00} & 57.29 \\
			$T_6$ & 0.31 & 5 & 33.13 & 0.000 & \textbf{19.67} & 0.000 & 1.000 & \textbf{16.00} & 5.286 \\
			$T_7$ & 0.28 & 5 & 31.43 & 33.00 & \textbf{80.67} & 55.60 & 23.00 & \textbf{78.00} & 40.43 \\
			$T_8$ & 0.39 & 7 & 29.28 & 0.000 & \textbf{0.222} & 0.000 & 0.000 & \textbf{1.000} & 0.000 \\
			$T_9$ & 0.33 & 5 & 31.18 & 0.000 & 0.000 & 0.000 & 0.000 & 0.000 & 0.000 \\
			
			\midrule
			
			\textit{Avg} & -- & -- & -- & 8.750 & \textbf{32.76} & 22.78 & 18.75 & \textbf{32.50} & 19.52 \\
			
			\bottomrule
			
		\end{tabular}
		\begin{tablenotes}[para]
			\item[$\ast$] \textit{Training track}
			\item[$\dagger$] \textit{Specular version of the training track}
		\end{tablenotes}
		
	\end{threeparttable}
\end{table}

\autoref{table:evaluation:rq3} shows the results of retraining the best driving models, i.e., $M_3$ and $M_\textit{last}$ (as they show similar performance in terms of radius and recoverability), using the respective boundary state pairs, both for \tool and \baseline. In particular, we show the \textit{Success Rate (\%)} on the evaluation tracks of the original and retrained models (Column~5--7 for $M_3$, and Columns~8--10 for $M_{\textit{last}}$). The success rates of $M^3_\textit{retr.}$ and $M^{\textit{last}}_\textit{retr.}$ for each driving track, are averaged across the 9 repetitions.

For both \tool and \baseline, we observe the absence of regressions of the retrained models on the training track $T_1$. In fact, the success rate on $T_1$ remains 100\%, both for $M^3_\textit{retr.}$ and $M^{\textit{last}}_\textit{retr.}$. For \tool, the success rate of the retrained models improves in all the evaluation tracks where the original model has a non-zero success rate (i.e., see $T_2$, $T_5$---$T_7$ for $M_3$ and $T_3$, $T_5$---$T_7$ for $M_\textit{last}$). On the other hand, we observe that the $M_\textit{last}$ model, when retrained with the boundary state pairs of \baseline (Column~10), has lower success rates w.r.t. the original model in $T_3$ (i.e., 0.143\% success rate vs 56\%) and $T_5$ (i.e., 57.29\% vs 70\%).

On average, the boundary state pairs of \tool consistently cause an improvement of the success rate, going from a minimum of 2.4$\times$ to a maximum of 4.7$\times$ for $M_3$, and from a minimum of 1.3$\times$ to a maximum of 16$\times$ for $M_{\textit{last}}$. For both models the difference between the success rates of the original vs the retrained model is statistically significant with a small $\hat{A}_{12}$ (if we were to remove the zeros for $T_3$, $T_4$, $T_8$ and $T_9$ for $M_3$ and $T_2$, $T_4$, $T_8$ and $T_9$ for $M_{\textit{last}}$, the effect size would be large). Indeed, some of the testing tracks are very challenging for $M_3$ and $M_\textit{last}$ (i.e., success rate of 0\%), and on them retraining with the boundary state pairs results in an improvement of the success rate only on a few of them (i.e., see $T_6$ for $M_3$ and $T_8$ for $M_{\textit{last}}$). 
Overall, the success rate of the retrained models is on average, across all evaluation tracks, 3.7$\times$ the success rate of the original model for $M_3$ and 1.7$\times$ for $M_\textit{last}$. 
Considering the boundary state pairs of \baseline, we observe an overall improvement of 2.6$\times$ for $M_3$, and only a slight improvement for $M_\textit{last}$ (i.e., 1.04$\times$ on average). However, for both driving models the difference between the success rates of the original vs the retrained models is not statistically significant, suggesting that the boundary state pairs of \baseline are not effective enough to significantly improve strong driving models.

\begin{tcolorbox}[colback=gray!15!white,colframe=black]
	\textbf{RQ\textsubscript{3} (Retraining)}: The boundary state pairs of our best driving models, when extracted by \tool, are effective at improving the model through retraining. On average, the success rate of the retrained model on a set of evaluation tracks is between 1.7$\times$ and 3.7$\times$ higher than the original success rate, with a maximum improvement of 16$\times$.
\end{tcolorbox}

\subsection{Threats to Validity} \label{sec:evaluation:threats}

\head{External Validity} Using one simulator and one ADS poses an external validity threat. To mitigate this issue we selected a widely used ADS from the testing literature~\cite{third-eye,stocco-mind,stocco-jsep,stocco-misbehaviour,deephyperion,deepjanus,sbft-2023,deepguard,adept} (i.e., Dave-2~\cite{nvidia-dave2}) and we considered it at different training levels. As driving simulator we selected the popular open-source driving simulator Donkey Car\textsuperscript{\texttrademark}, also used in previous studies in ADS testing~\cite{stocco-mind,testing-drl-agents}. Both the simulator and the ADS we selected represent the state-of-the-art in ADS testing as a recent survey suggests~\cite{survey-ads}. 
Another external validity threat is the limited number of evaluation tracks we used to measure the improvement of the retrained driving model. We addressed this threat by generating a diverse set of evaluation tracks where the original driving model displayed a wide range of success rates, i.e., ranging from high (70\%) to low (0\%).

Focusing on lane-keeping task, also poses an external validity threat. Although our approach might not generalize to other tasks (such as pedestrian avoidance~\cite{borg} or collision avoidance~\cite{morlot,calo-generating}), it only requires the formalization of the concept of state as well as the definition of the closeness and validity constraints. Our overall approach of searching for boundary state pairs and improving the driving model is generally applicable to any driving task, although the definition of the closeness and validity constraints needs to be specific to the state variables at hand.

\head{Internal Validity} We compared all the driving models under identical  settings of the hyperparameters of the search algorithm and
we evaluated the improvement of the retrained model on the same evaluation tracks. 

\head{Conclusion Validity} We executed the search process multiple times with different random seeds to account for the randomness of the search algorithm. We also ran each driving simulation under the same conditions multiple times to both assess the reliability of a boundary state pair and to measure the success rate. We used rigorous statistical tests to draw our conclusions.

\head{Reproducibility} The source code, the driving simulator, and the test objects are publicly available in our replication package~\cite{replication-package}.

	\section{Related Work} \label{sec:related-work}

\head{Simulation-based Testing} Simulation-based testing is a prominent approach in the literature to test the reliability of autonomous driving systems (ADSs)~\cite{survey-ads}. Several techniques have been proposed by researchers. \textsc{AsFault} combines procedural content generation and search-based algorithms to generate virtual roads. Abdessalem et al.~\cite{icse18-abdessalem} use decision trees to guide the search towards the critical features of the environment. Calò et al.~\cite{calo-generating} introduce the concept of avoidable collisions and proposed two search-based algorithms to generate them. Riccio et al. propose \textsc{DeepJanus}~\cite{deepjanus} that use multi-objective optimization to generate challenging and diverse virtual roads; moreover, they propose \textsc{DeepHyperion}~\cite{deephyperion,deephyperion-tosem} that outputs an explanatory map of the road features that cause a misbehavior. \textsc{Samota}~\cite{samota} and \textsc{Indago}~\cite{testing-drl-agents}, respectively for Deep Learning (DL) and Reinforcement Learning (RL) systems, use a surrogate model to predict the output of the driving simulator in order to efficiently generate critical scenarios. Similarly, Giamattei et al.~\cite{cart} propose \textsc{Cart} that builds a causal model of the environment and queries the model to execute only the most promising test cases in the simulator. Recently, \textsc{Morlot}~\cite{morlot} and \textsc{DeepCollision}~\cite{deepcollision} use reinforcement learning (RL) to manipulate the large input space of the driving simulation, while Doreste et al.~\cite{adversarial-rl} use RL to train an intelligent testing agent. 
Our approach \tool differs from the above in three distinct ways: (1)~\tool starts from a failure-free driving scenario to extract hidden and challenging driving conditions; (2)~\tool mutates the driving conditions of the ego vehicle instead of acting on the environment; (3)~\tool uses the boundary state pairs associated with challenging driving conditions to improve the ADS under test. Such differences make \tool (i.e., a \textit{driving-condition-mutating} approach) and the \textit{scenario-mutating} approaches described above, complementary. Indeed, scenario-mutating approaches can explore challenging driving scenarios within the scenario space, while driving-condition-mutating approaches can extract challenging driving conditions within those scenarios.

\head{Boundary Input Generation} Instead of generating individual driving scenarios that induce a failure of the ADS under test, researchers proposed search-based approaches to find boundary inputs~\cite{pezze-software}, i.e., inputs that trigger different behaviors of the ADS under test. For instance, Mullins et al.~\cite{mullins-frontier} use an adaptive search algorithm to discover performance boundaries of the ADS. Tuncali et al.~\cite{tuncali-frontier} employ an approach called rapidly-exploring random trees to generate a pair of configurations where a collision is avoidable/unavoidable. Riccio et al.~\cite{deepjanus} use a multi-objective search algorithm to generate a road shape where the ADS starts to misbehave. 
In the approach by Biagiola et al.~\cite{plasticity-rl}, a test case is a pair of environment configurations and the objective is to find boundary pairs such that in one of them the RL agent under test can adapt to the new environment, while in the other it cannot. 
While such approaches search for boundary pairs of the environment, our approach focuses on the boundary pairs of the vehicle the driving model controls.

Similarly to us, Tappler et al.~\cite{tappler-search} define a boundary state as a state  that precedes another non-terminal state in which the RL agent under test misbehaves. However, their algorithm to search for boundary states relies on symbolic reasoning and symbolic exploration of the execution tree, which does not scale to a complex context such as ADS testing, for which we designed a novel, ad-hoc state exploration algorithm.

	\section{Conclusion and Future Work} \label{sec:conclusion}

\tool extracts challenging driving conditions in a failure-free driving scenario by mutating the initial state of the ADS. Experimental results show that boundary state pairs exist even for high-quality and well-trained driving models.
By retraining the ADS under test with examples collected from its boundary state pairs, we significantly improved the success rate of the retrained model on a separate set of evaluation tracks. In our future work, we plan to extend our study to multiple driving simulators to extend the generalizability of our approach and experiment with alternative search algorithms to look for boundary state pairs. We also plan to investigate other safety-critical case studies other than autonomous driving (e.g., unmanned aerial vehicles~\cite{aerialist-icst}), to study the failure boundary of other learning-based systems.

	\section{Acknowledgements}
This work was partially supported by the H2020 project PRECRIME, funded under the ERC Advanced Grant 2017 Program (ERC Grant Agreement n. 787703).

	\balance
	\bibliographystyle{ieeetr}
	\bibliography{paper}
	
\end{document}